\documentclass[twocolumn,showpacs,preprintnumbers,amsmath,amssymb]{revtex4}
\usepackage{graphicx}% Include figure files
\usepackage{dcolumn}% Align table columns on decimal point
\usepackage{bm}% bold math
\begin{document}

\title{{\Large Feynman and Squeezed States}}

\author{Y. S. Kim}
\affiliation{Center for Fundamental Physics. University of
  Maryland, College Park, Maryland 20742. U.S.A.}

\begin{abstract}

In 1971, Feynman {\em et al.} published a paper on hadronic mass
spectra and transition rates based on the quark model.  Their
starting point was a Lorentz-invariant differential equation.
This equation can be separated into a Klein-Gordon equation for
the free-moving hadron and a harmonic oscillator equation for the
quarks inside the hadron.  However, their solution of the oscillator
equation is not consistent with the existing rules of quantum
mechanics and special relativity.  On the other hand, their partial
differential equation has many other solutions depending on boundary
conditions.  It is noted that there is a Lorentz-covariant set of
solutions totally consistent with quantum mechanics and special
relativity.  This set constitutes a representation of the Poincar\'e
group which dictates the fundamental space-time symmetry of particles
in the Lorentz-covariant world.  It is then shown that the same set
of solutions can be used as the mathematical basis for two-photon
coherent states or squeezed states in quantum optics.  It is thus
possible to transmit the physics of squeezed states into the hadronic
world.  While the time-like separation is the most puzzling problem
in the covariant oscillator regime, this variable can be interpreted
like the unobserved photon in the two-mode squeezed state which leads
to an entropy increase.
\end{abstract}

\pacs{11.30.Cp, 12.39.Ki, 42.50.-p}

\maketitle

\section{Introduction}\label{intro}
Since Einstein's formulation of special relativity in 1905, the
most important development in physics is the formulation of
quantum mechanics resulting in Heisenberg's uncertainty principle.

For solving practical problems, the Schr\"odinger wave equation
is commonly used.  For scattering problems, we use running-wave
solutions.  For bound states, we obtain standing-wave solutions
with their boundary conditions.  Indeed, this localization boundary
condition leads to discrete energy levels.

For scattering problems, we now have Lorentz-covariant quantum
field theory with its scattering matrix formalism and Feynman
diagrams.  Since quantum field theory was so successful that there
had been attempts in the past to understand bound-state problems
using the S matrix method.  However, it was noted from the
calculation of the neutron-proton mass difference from Dashen and
Frautchi that the S-matrix method does not guarantee the localization
of the bound-state wave functions~\cite{dyson65, kim66}.

While the localization boundary condition is the main problem for
the bound state, the question is whether this issue is covariant
under Lorentz transformations.  We know the hydrogen has its
localized wave function, but how would this appear to the observer
on a train?  One way to reduce this difficulty is to study harmonic
oscillators, because the oscillator system has its built-in boundary
condition.  For this reason, there had been many attempts in the
past to make the harmonic oscillator Lorentz-covariant.

In order to understand the hadronic mass spectra and hadronic
transition rates in the quark model, Feynman {\em et al.} in 1971
published a paper containing the following Lorentz-invariant
differential equation~\cite{fkr71}.

\begin{widetext}

\begin{equation}\label{diff11}
\left\{-\frac{1}{2}\left[\left(\frac{\partial}
                                {\partial x^a_{\mu}}\right)^2 +
\left(\frac{\partial}{\partial x^b_{\mu}}\right)^2 \right]  +
\frac{1}{16}\left(x^a_{\mu} -x^b_{\mu}\right)^2 + m_0^2\right\}
\phi\left(x^a_{\mu}, x^b_{\mu}\right) = 0 ,
\end{equation}
for a hadron consisting of two quarks bound-together harmonic
oscillator potential.  The space-time quark coordinates are
$x^a_{\mu}$ and $x^b_{\mu}$.  They then wrote down the equation

\end{widetext}

They wrote down the hadronic and quark separation coordinates
as
\begin{eqnarray}\label{coord}
&{}& X_\mu = \frac{1}{2}\left(x^a_{\mu} + x^b_{\mu} \right),
                    \nonumber \\[1ex]
&{}& x_\mu = \frac{1}{2\sqrt{2}}\left(x^a_{\mu} - x^b_{\mu}\right) ,
\end{eqnarray}
respectively, and $\phi\left(x^a_{\mu},x^b_{\mu}\right)$ as
\begin{equation}
\phi\left(x^a_{\mu},x^b_{\mu}\right) = f\left(X_{\mu}\right)
     \psi\left(x_{\mu}\right) .
\end{equation}
Then the differential equation can be separated into the following
two equations.
\begin{equation}\label{diff22}
\left\{\left(\frac{\partial}{\partial X_{\mu}}\right)^2 + m_0^2 +
(\lambda + 1) \right\}
f\left(X_{\mu}\right) = 0 ,
\end{equation}
for the hadronic coordinate, and
\begin{equation}\label{diff33}
\frac{1}{2} \left\{-\left(\frac{\partial}{\partial x_{\mu}}\right)^2
  + x_{\mu}^2 \right\} \psi\left(x_{\mu}\right) = (\lambda + 1)
 \psi \left(x_{\mu}\right) ,
\end{equation}
for the coordinate of quark separation inside the hadron.

The differential equation of Eq.(\ref{diff22}) is a Klein-Gordon
equation for the hadronic coordinate.  The Klein-Gordon equation
is Lorentz-invariant and is the starting point for quantum field
theory for scattering processes with Feynman diagrams.  This
aspect of physics is well known.  In the present case, the
solution takes the form
\begin{equation}
f(X) = \exp{ \left( \pm iP\cdot X \right) } ,
\end{equation}
with
\begin{equation}
 - P^2 = m_0^2 + (\lambda + 1) .
 \end{equation}
We are using here the space-favored metric where $P^2 =
\left(P_x^2 + P_y^2 + P_z^2 - E^2\right)$.  The hadronic mass
is thus determined from $m_0$ and $\lambda$.  The $\lambda$
parameter is determined from the oscillator equation of
Eq.(\ref{diff33}) for the internal space-time coordinate.   The
internal quark motion determined the hadronic mass, according to
Feynman {\em et al.}

Indeed, the differential equation of Eq.(\ref{diff11}) contains
the scattering-state equation for the hadron, and the bound-state
equation for the quarks inside the hadron.  The differential
equation of Eq.(\ref{diff33}) is also a Lorentz-invariant equation.
The problem is that the set of solutions given by Feynman
{\em et al.} in their 1971 paper is not consistent with the
existing rules of physics.  This is the reason why this paper
is not well known.

However, this does not exclude other sets of solutions.  The
solutions can take different forms depending on the separable
coordinate systems with their boundary conditions.  Indeed,
there is a set of oscillator solution that can constitute a
representation of the Poincar\'e group, particularly that of
Wigner's little group which dictates the internal space-time
symmetry of the particles in the Lorentz-covariant
world~\cite{wig39,kno79}.  We choose to call this set of
solutions the Poincar\'e set.

If we ignore the time-like variable in Eq.(\ref{diff33}), it
is the Schr\"odinger-type equation for the three-dimensional
harmonic oscillator.  If we ignore it, the equation loses its
Lorentz invariance.  The problem is how to deal with the
time-separation variable, while it is not even mentioned in
the present form of quantum mechanics.  If we believe in
Einstein, this variable exists wherever there is a spacial
separation like the Bohr radius.  However, we pretend know
about it in the present form of quantum mechanics.

In this report, we give an interpretation to this variable
based on the lessons we learn from quantum optics.  For this
purpose, we show first that the Poincar\'e set for the
covariant harmonic oscillator can be used for mathematical
basis for two-photon coherent states or squeezed
states~\cite{yuen76}.  In other words, the squeezed states
can be constructed from the Lorentz-invariant differential
equation of Eq.(\ref{diff33}).

We then establish that the longitudinal and time-like
excitations in the covariant harmonic oscillator system can
be translated into the two-photon coherent state.  We know
what happens when one of the two photons is not observed.
The result is an increase in entropy~\cite{ekn89}.  We can
then go back to the covariant oscillator and give a similar
interpretation to the time-separation variable which is
not observed in the present form of quantum mechanics,

In Sec.~\ref{lorentz}, we introduce the set of solutions of
Eq.(\ref{diff33}) which constitutes a representation of the
Poincar\'group.  We then study the space-time geometry of
its Lorentz covariance.  In Sec.~\ref{sqz}, it is shown that
the oscillator differential equation of Eq.(\ref{diff33}) can
serve as the starting equation for squeezed states in quantum
optics, and also that this Poincar\'e set serves as the
mathematical basis for the two-photon coherent state.  In
Sec.~\ref{restof}, we give a physical interpretation to the
time-separation variable in terms of Feynman's rest of the
universe, which has a concrete physical interpretation in
quantum optics.

\section{Lorentz Boosts as Squeeze Transformations}\label{lorentz}

In 1979, Kim, Noz, and Oh published a paper on representations
of the Poincar\'e group using a set of solutions of the oscillator
equation of Eq.(\ref{diff33})~\cite{kno79}.  Later in 1986, Kim
and Noz in their book~\cite{knp86} noted that this set corresponds
to a representation of Wigner's $O(3)$-like little group for
massive particles.  If a particle has a non-zero mass, there is
a Lorentz frame in which the particle is at rest.  Wigner's little
group then becomes that of the three-dimensional rotation group,
which is very familiar to us.

The Lorentz-covariant solution of the Lorentz-invariant differential
equation contains both space-like and time-like wave components,
but we can keep the time-like component to its ground state.  The
wave function thus retains the $O(3)$-like symmetry.  The solution
takes the form
\begin{equation}\label{sol00}
\psi(x,y,z,t) =
  \left\{\left(\frac{1}{\pi}\right)^{1/4}
     \exp{\left(\frac{-t^2}{2}\right)}\right\}\psi(x,y,x) .
\end{equation}

As for the spatial part of the differential equation, it is the
equation for the three-dimensional oscillator.  We can solve this
equation with both the Cartesian and spherical coordinate systems.
If we use the spherical system with $(r, \theta, \phi)$ as the
variables, the solution should take the form
\begin{equation}\label{sol22}
\psi(x,y,z) =  R_{\lambda,\ell}(r)Y_{\ell,m}(\theta,\phi)
\exp{\left\{ -\left(\frac{x^2 + y^2 +z^2}{2}\right)\right\}} ,
\end{equation}
where $Y_{\ell,m}(\theta,\phi)$ is the spherical harmonics, and
$R_{\lambda,\ell}(r)$ is the normalized radial wave function
with $r = \sqrt{x^2 + y^2 + z^2}.$  The $\lambda$ and $\ell$
parameters specify the mass and the internal spin of the hadron
respectively, as required by Wigner's representation
theory~\cite{wig39,knp86}.

If we use the Cartesian coordinate systems, and the solution
can be written as

\begin{widetext}

\begin{equation}\label{sol33}
\psi(x,y,z)=
  \left[\frac{1}{\pi\sqrt{\pi} 2^{(a+b+n)}a!b!n!}\right]^{1/2}
           H_a(x) H_b(y) H_n(z)
  \exp{\left\{-\left(\frac{x^2 + y^2 +z^2}{2}\right)\right\}} ,
\end{equation}
where $H_n(z)$ is the Hermite polynomial of $z$.  Since the
three-dimensional oscillator system is separable in both the
spherical and Cartesian coordinate systems, the wave function
of Eq.(\ref{sol22}) can be written as a linear combination of
the solutions given in Eq.(\ref{sol33}), with
$\lambda = a + b + n.$

When we boost this solution along the $z$ direction, the
Cartesian form of Eq.(\ref{sol33}) is more convenient.  Since
the transverse $x$ and $y$ coordinates are not affected by this
transformation, we can separate out these variables in the
oscillator differential equation of Eq.(\ref{diff33}), and
consider the differential equation
\begin{equation}\label{diff44}
\frac{1}{2} \left\{\left[-\left(\frac{\partial}{\partial z}\right)^2
 + z^2 \right]
 -\left[-\left(\frac{\partial}{\partial t}\right)^2 +
 t^2\right]\right\}\psi(z,t) = n \psi(z,t) .
\end{equation}

\end{widetext}

This differential equation remains invariant under the Lorentz
boost
\begin{eqnarray}\label{boostm}
&{}& z \rightarrow (\cosh\eta)z + (\sinh\eta)t , \nonumber \\[1ex]
&{}& t \rightarrow (\sinh\eta)z + (\cosh\eta)t .
\end{eqnarray}
with
\begin{equation}
e^{\eta} = \sqrt{\frac{1 + \beta}{1 - \beta}} ,
\end{equation}
where $\beta$ is the velocity parameter $v/c$.

If we suppress the excitations along the $t$ coordinate, the
normalized solution of this differential equation is
\begin{equation}\label{sol44}
\psi(z,t) =  \left(\frac{1}{\pi 2^{n}n!} \right)^{1/2}
 H_n(z)\exp{\left\{-\left(\frac{z^2 +t^2}{2}\right)\right\}} .
\end{equation} \\[1ex]
If we boost the hadron along the $z$ direction according to
Eq.(\ref{boostm}), the coordinate variables $z$ and $t$ should
be replaced respectively by $[(\cosh\eta)z - (\sinh\eta)t]$
and $[(\cosh\eta)t - (\sinh\eta)z]$ respectively, and the
expression becomes uncontrollable.

In his 1949 paper~\cite{dir49}, Dirac introduced his light-cone
variables defined as
\begin{equation}\label{lcvari}
u = \frac{z + t}{\sqrt{2}} , \qquad v = \frac{z - t}{\sqrt{2}} .
\end{equation}
Then the boost transformation of Eq.(\ref{boostm}) takes the form
\begin{equation}\label{lorensq}
u \rightarrow e^{\eta } u , \qquad v \rightarrow e^{-\eta } v .
\end{equation}
The $u$ variable becomes expanded while the $v$ variable becomes
contracted.  Their product
\begin{equation}
uv = \frac{1}{2}(z + t)(z - t) = \frac{1}{2}\left(z^2 - t^2\right)
\end{equation}\\[1ex]
remains invariant.  Indeed, in Dirac's picture, the Lorentz boost is a
squeeze transformation.

In terms of these light-cone variables, the ground-state wave function
becomes

\begin{widetext}

\begin{equation}\label{cwf11}
 \psi_{0}^{n}(x,t) = \left[\frac{1}{\pi n! 2^{n}} \right]^{1/2}
       H_{n}\left(\frac{u + v}{\sqrt{2}}\right)
       \exp{\left\{-\left(\frac{u^{2} + v^{2}}{2}\right)\right\}} ,
\end{equation}
and the excited-state wave function of Eq.(\ref{sol44}) takes
the form
\begin{equation}\label{cwf22}
 \psi_{\eta}^{n}(x,t) = \left[\frac{1}{\pi n! 2^{n}} \right]^{1/2}
       H_{n}\left(\frac{e^{-\eta}u +  e^{\eta} v}{\sqrt{2}}\right)
   \exp{\left\{-\left(\frac{e^{-2\eta}u^{2} + e^{2\eta}v^{2}}{2}
     \right)\right\}} ,
\end{equation}
for the moving hadron.  If we use the $x$ and $t$ variables,
\begin{equation}\label{cwf22a}
\psi_{\eta}^{n}(x,t) = \left[\frac{1}{\pi n! 2^{n}} \right]^{1/2}
       H_{n}\left(\frac{e^{-\eta}(z + t) +  e^{\eta}(z - t)}{2}\right)
 \exp{\left\{-\left[\frac{e^{-2\eta}(z + t)^{2}
       + e^{2\eta}(z - t)^{2}}{2} \right]\right\}} ,
\end{equation}

For the ground state with $n = 0,$ the wave function is a Gaussian
function
\begin{equation}
 \psi_{\eta}^{0}(x,t) = \left[\frac{1}{\pi} \right]^{1/2}
   \exp{\left\{-\left(\frac{e^{-2\eta}u^{2} + e^{2\eta}v^{2}}{2}
     \right)\right\}} .
\end{equation}
In terms of the $z, t$ variables,
\begin{equation}
 \psi_{\eta}^{0}(x,t) = \left[\frac{1}{\pi} \right]^{1/2}
   \exp{\left\{-\left[\frac{e^{-2\eta}(z + t)^{2} +
   e^{2\eta}(z - t)^{2}}{4} \right] \right\}} .
\end{equation}

\end{widetext}

%-----------------------------------------------------------------
\begin{figure}%[thb]
\centerline{\includegraphics[scale=0.38]{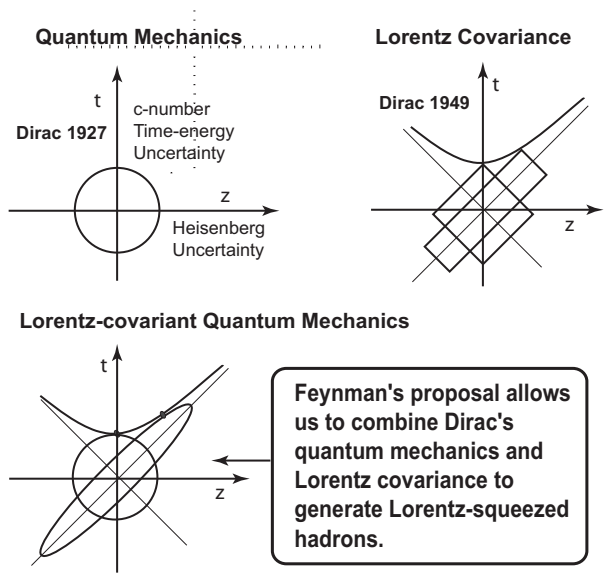}}
\vspace{2mm}
\caption{Lorentz-squeezed hadrons.  Feynman's proposal leads us
to combine Dirac's quantum mechanics with c-number time-like
excitation~\cite{dir27}, and his light-cone representation of
Lorentz boosts~\cite{dir45}.  Dirac also considered using
harmonic oscillators to combine quantum mechanics and special
relativity~\cite{dir45,dir63}.  It is not difficult to obtain
the third figure by combining the first two.  The Lorentz boost
is a squeeze transformation.}\label{diracqm33}
\end{figure}
%------------------------------------------------------------------

This Gaussian factor determines the space-time localization
property of all excited-state wave functions, and its space-time
localization property is illustrated in terms of the circle
and ellipse in Fig.~\ref{diracqm33}. According to this figure,
the Lorentz boost is a squeeze transformation.  This figure
combines Dirac's four papers aimed at combining quantum mechanics
with special relativity~\cite{dir49,dir27,dir45,dir63}.

It is important to note that this Lorentz squeeze property has
been experimentally verified in various observations in
high-energy physics, including Feynman's parton
picture~\cite{knp86,fey69a,kim89}.

\section{Squeezed States}\label{sqz}
Let us start with the Hamiltonian of the form
\begin{equation}\label{hamil+}
H_{+} = \frac{1}{2}
 \left\{\left[-\left(\frac{\partial}{\partial x_{1}}\right)^2
 + x_{1}^2 \right]
 + \left[-\left(\frac{\partial}{\partial x_{2}}\right)^2 +
 x_{2}^2\right]\right\} ,
\end{equation}
and the differential equation
\begin{equation}\label{diff88}
 H_{+} \psi\left(x_{1},x_{2}\right) = \left(n_{1} + n_{2} + 1\right)
  \psi\left(x_{1},x_{2}\right) .
\end{equation}
This is the Schr\"odinger equation for the two-dimensional harmonic
oscillator.  This differential equation is separable in the $x_1$
and $x_2$ variables, and the wave function can be written as
\begin{equation}
  \psi\left(x_{1},x_{2}\right) = \chi_{n_1}\left(x_{1}\right)
       \chi_{n_2}\left(x_{2}\right) ,
\end{equation}\\[1ex]
where $\chi_{n}(x)$ is the $n$-th excited-state oscillator wave
function which takes the form
\begin{equation}\label{chi11}
     \chi_n (x) = \left[\frac{1}{\sqrt{\pi}2^n n!}\right]^{1/2}
              H_n(x) \exp{\left(\frac{-x^2}{2}\right)} .
\end{equation}
Thus

\begin{widetext}

\begin{equation} \label{cwf66}
  \psi\left(x_{1},x_{2}\right) =
      \left[\frac{1}{\pi 2^{(n_1 + n_2)} (n_1 + n_2)!}\right]^{1/2}
        H_{n_1}\left(x_1\right) H_{n_2}\left(x_2\right)
         \exp{\left\{- \frac{1}{2}\left(x_1^2 + x_2^2 \right)\right\}} .
\end{equation}
If the system is in the ground state with $n_{1} = n_{2} = 0$, this
wave function becomes
\begin{equation}\label{cwf99}
 \psi\left(x_{1},x_{2}\right) = \left[\frac{1}{\pi}\right]^{1/2}
        \exp{\left\{- \frac{1}{2}\left(x_1^2 + x_2^2 \right)\right\}} .
\end{equation}

If the $x_2$ coordinate alone is in its ground state, the wave
function becomes
\begin{equation}\label{cwf88}
   \psi\left(x_{1},x_{2}\right) =
               \left[\frac{1}{\pi 2^n n!}\right]^{1/2}
          H_{n} \left(x_1\right)
        \exp{\left\{- \frac{1}{2}\left(x_1^2 + x_2^2 \right)\right\}} ,
\end{equation}
with $n = n_1$.
In order to squeeze this wave function, we introduce first the normal
coordinates
\begin{equation}\label{normal11}
y_1 = \frac{1}{\sqrt{2}} \left( x_{1} + x_{2}\right), \qquad
y_2 = \frac{1}{\sqrt{2}} \left( x_{1} - x_{2}\right) .
\end{equation}
In terms of these variables, the wave function of Eq.(\ref{cwf99}) can
be written as
\begin{equation}\label{cwf55}
 \psi_{0}^{n}(x_1, x_2) = \left[ \frac{1}{\pi n! 2^{n}} \right]^{1/2}
       H_{n}\left(\frac{y_1 + y_2}{\sqrt{2}}\right)
       \exp{\left\{-\left(\frac{y_1^{2} + y_2^{2}}{2}\right)\right\}} ,
\end{equation}

Let us next squeeze the system by making the following coordinate
transformation.
\begin{equation}
 y_{1} \rightarrow e^{\eta} y_1,  \qquad
 y_{2} \rightarrow e^{-\eta} y_1 .
\end{equation}
This transformation is equivalent to
\begin{equation}\label{boostm22}
x_{1} \rightarrow \left[(\cosh\eta)x_1 +
             (\sinh\eta)x_2 \right] , \qquad
x_{2} \rightarrow \left[(\sinh\eta)x_1 +
             (\cosh\eta)x_2 \right] ,
\end{equation}
like the Lorentz boost given in Eq.(\ref{boostm}).

The wave function then becomes
\begin{equation}\label{cwf11a}
\psi_{\eta}^{n}\left(x_{1}, x_{2}\right) =
 \left[\frac{1}{\pi n! 2^{n}} \right]^{1/2}
  H_{n}\left(\frac{e^{-\eta}y_{1} +
       e^{\eta}y_{2}}{\sqrt{2}}\right)
     \exp{\left\{-\left[\frac{e^{-2\eta}y_1^{2} +
           e^{2\eta}y_2^{2}}{2}\right]\right\}} ,
\end{equation}
If we use the $x_1$ and $x_2$ variables, this expression becomes
\begin{equation}\label{cwf11b}
\psi_{\eta}^{n}\left(x_{1}, x_{2}\right) =
 \left[\frac{1}{\pi n! 2^{n}} \right]^{1/2}
  H_{n}\left(\frac{e^{-\eta}  \left(x_{1} + x_2\right)
  e^{\eta} \left(x_{1} - x_2\right)}{2}\right)
     \exp{\left\{-\left[\frac{e^{-2\eta} \left(x_{1} +
     x_2\right) ^{2} +  e^{2\eta} \left(x_{1} -
     x_{2}\right)^{2}}{2}\right]\right\}} .
\end{equation}
This transformed wave function does not satisfy the eigenvalue
equation of Eq.(\ref{diff88}).  It is a linear combinations of the
eigen solutions $\chi\left(x_1\right)$ and $\chi\left(x_1\right)$
defined in Eq.(\ref{chi11}).  The linear expansion takes the
form~\cite{knp86,knp91}
\begin{equation}\label{cwf33}
    \psi_{\eta}^{n}\left(x_1,x_2\right) =
      \left(\frac{1}{\cosh\eta}\right)^{(n+1)}
     \sum_{k} \left[\frac{(n+k)!}{n!k!}\right]^{1/2}
  (\tanh\eta)^{k}\chi_{n+k}\left(x_1\right)\chi_{k}\left(x_2\right) ,
\end{equation}
In quantum optics, the eigen functions $\chi_{n+k}\left(x_1\right)$
and $\chi_{k}\left(x_1\right)$ correspond to the $(n + k)$-photon
state of the first photon and $k$-photon state of the second photon
respectively.

If $n = 0$, it becomes the squeezed ground state or vacuum state,
and the resulting wave function is
\begin{equation}\label{cwf330}
    \psi_{\eta}^{0}\left(x_1,x_2\right) =
    \left(\frac{1}{\cosh\eta}\right)
      \sum_{k} (\tanh\eta)^{k}\chi_{n+k}
      \left(x_1\right)\chi_{n}\left(x_2\right) ,
\end{equation}
In the literature, this squeezed ground state is known
as the squeezed vacuum state~\cite{yuen76}, while the
expansion of Eq.(\ref{cwf33}) is for the squeezed $n$-photon
state~\cite{knp91}.

While these wave functions do not satisfy the eigenvalue
equation with the Hamiltonian of Eq.(\ref{hamil+}), they
satisfy the eigenvalue equation with the Hamiltonian $H_{-}$,
where
\begin{equation}\label{hamil-}
H_{-} = \frac{1}{2} \left\{\left[-\left(\frac{\partial}
   {\partial x_{1}}\right)^2  + x_{1}^2 \right]
 - \left[-\left(\frac{\partial}{\partial x_{2}}\right)^2 +
 x_{2}^2\right]\right\} .
\end{equation}
If the $x_2$ coordinate is in its ground state,
\begin{equation}\label{diff99}
 H_{-} \psi\left(x_{1},x_{2}\right) =
                n \psi\left(x_{1},x_{2}\right) .
\end{equation}

\end{widetext}

If we replace the notations $x_{1}$ and $x_{2}$ by $z$ and $t$
respectively, this Hamiltonian becomes that of Eq.(\ref{diff44}).
Thus, the oscillator equation of Eq.(\ref{diff44}) generates
a set of solutions which forms the basis for the squeezed states.

The $t$ variable is the time-separation variable, and is not
the time variable appearing in the time-dependent Schr\"odinger
equation.  We shall discuss this variable in detail in
Sec.~\ref{restof}.

The differential equation of Eq.(\ref{diff44}) was proposed by
Feynman {\it et al.} in 1971~\cite{fkr71}.  Even though they
were not able to provide physically meaningful solutions to their
own equation, it is gratifying to note that there is at least
one set of solutions which can explain many aspects of physics,
including squeezed states in quantum optics as well as the
basic observable effects in high-energy hadronic physics.

\section{Feynman's Rest of the Universe}\label{restof}

In Sec.~\ref{lorentz}, the time-separation variable played a major
role in making the oscillator system Lorentz-covariant.  It should
exist wherever the space separation exists.
The Bohr radius is the measure of the separation between the proton
and electron in the hydrogen atom.  If this atom moves, the radius
picks up the time separation, according to Einstein~\cite{kn06aip}.

On the other hand, the present form of quantum mechanics does not
include this time-separation variable.  The best way we can do at
the present time is to treat this time-separation as a variable in
Feynman's rest of the universe~\cite{hkn99ajp}.  In his book on
statistical mechanics~\cite{fey72}, Feynman states

\begin{quote}
{\it When we solve a quantum-mechanical problem, what we really do
is divide the universe into two parts - the system in which we are
interested and the rest of the universe.  We then usually act as if
the system in which we are interested comprised the entire universe.
To motivate the use of density matrices, let us see what happens
when we include the part of the universe outside the system.}
\end{quote}

The failure to include what happens outside the system results in
an increase of entropy.  The entropy is a measure of our ignorance
and is computed from the density matrix~\cite{neu32}.  The density
matrix is needed when the experimental procedure does not analyze
all relevant variables to the maximum extent consistent with
quantum mechanics~\cite{fano57}.  If we do not take into account
the time-separation variable, the result is therefore an increase
in entropy~\cite{kiwi90pl,kim07}.

It is gratifying to note that the two-mode coherent state in quantum
optics shares the same mathematical basis as the covariant harmonic
oscillator.  In the two-mode squeezed state, both photons are
observable, but the physics survives and becomes even more
interesting if one of them is not observed~\cite{ekn89}.

%-------------------------------------------------------------------
\begin{figure}%[thb]
\vspace{5mm}
\centerline{\includegraphics[scale=0.26]{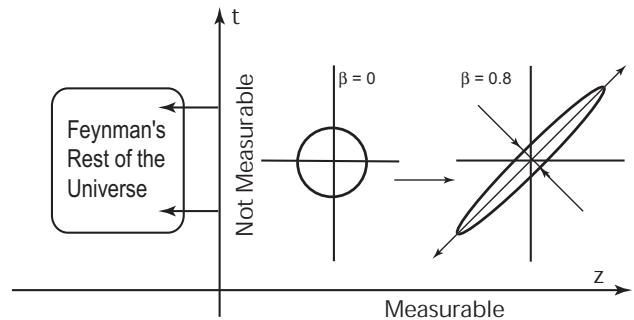}}
\vspace{2mm}
\caption{Localization property in the $zt$ plane.  When the hadron
is at rest, the Gaussian form is concentrated within a circular
region specified by $(z + t)^2 + (z - t)^2  = 1.$  As the hadron
gains speed, the region becomes squeezed to $e^{-2\eta}(z + t)^2
+ e^{2\eta}(z - t)^2 = 1.$  Since it is not possible to make
measurements along the $t$ direction, we have to deal with the
information less than complete, and to resort to
entropy.}\label{restof33}
\end{figure}
%----------------------------------------------------------------------

In the covariant oscillator formalism, these two photons are
translated into longitudinal and time-like excitations in the
hadronic system.  If the hadron is at rest, there are no time-like
excitations.  On the other hand, if the hadron moves, there are
time-like excitations to the observer at rest.  But this observer
is not able to detect it.  Indeed, these time-like oscillations are
in Feynman's rest of the universe.

Let us carry out a concrete mathematics using the density matrix
formalism.  From the covariant oscillator wave functions defined
Sec.~\ref{lorentz}, the pure-state density matrix is
\begin{equation}\label{den11}
  \rho_\eta^{n}(z,t;z',t') = \psi_\eta^{n}(z,t) \psi_\eta^{n} (z',t') ,
\end{equation}
which satisfies the condition $\rho^2  = \rho: $
\begin{equation}
  \rho_\eta^{n}(z,t;x',t') = \int \rho_\eta^{n}(z,t;x",t")
   \rho_\eta^{n}(z",t";z',t') dz"dt" .
\end{equation}

However, in the present form of quantum mechanics, it is not possible
to take into account the time separation variables.  Thus, we have to
take the trace of the matrix with respect to the t variable.  Then
the resulting density matrix is~\cite{kiwi90pl}

\begin{widetext}

\begin{eqnarray}\label{den22}
&{}& \rho_\eta^{n}(z,z') = \int \psi_\eta^{n}(z,t)
                            \psi_\eta^{n}(z',t) dt \nonumber \\[2ex]
&{}& \hspace{15mm}  = \left(\frac{1}{\cosh\eta}\right)^{2(n + 1)}
     \sum_{k} \frac{(n+k)!}{n!k!}
     (\tanh\eta)^{2k}\psi_{n+k}(z)\psi^*_{k+n}(z') .
\end{eqnarray}
The trace of this density matrix is one, but the trace of $\rho^2$
is less than one, as we can see from the following formula.
\begin{eqnarray}
&{}& Tr\left(\rho^2\right) = \int \rho_\eta^{n}(z,z')
                     \rho_\eta^{n}(z',z) dz dz' \nonumber \\[2ex]
  &{}& \hspace{15mm} = \left(\frac{1}{\cosh\eta}\right)^{4(n + 1)}
  \sum_{k} \left[\frac{(n+k)!}{n!k!}\right]^2 (\tanh\eta)^{4k} .
\end{eqnarray}
which is less than one.  This is due to the fact that we do not know
how to deal with the time-like separation in the present formulation
of quantum mechanics.  Our knowledge is less than complete.

The standard way to measure this ignorance is to calculate the
entropy defined as
\begin{equation}
          S = - Tr\left(\rho \ln(\rho)\right)  .
\end{equation}

If we can measure the distribution along the time-like direction
and use the pure-state density matrix given in Eq.(\ref{den11}),
the entropy is zero.  However, if we do not know how to deal
with the distribution along $t$, then we should use the density
matrix of Eq.(\ref{den22}) to calculate the entropy, and the
result is~\cite{kiwi90pl}
\begin{equation}
  S = 2(n + 1)\left\{(\cosh\eta)^2 \ln(\cosh\eta) -
              (\sinh\eta)^2 \ln(\sinh\eta)\right\}
 - \left(\frac{1}{\cosh\eta}\right)^{2(n + 1)}
  \sum_{k} \frac{(n+k)!}{n!k!}\ln\left[\frac{(n+k)!}{n!k!}\right]
               (\tanh\eta)^{2k} .
\end{equation}
In terms of the velocity $v$ of the hadron,
\begin{equation}
 S = -(n + 1)\left\{\ln\left[1 - \left(\frac{v}{c}\right)^2\right]
      + \frac{(v/c)^2 \ln(v/c)^2}{1 - (v/c)^2}  \right\}
 - \left[1 - \left(\frac{1}{v}\right)^2\right]
  \sum_{k} \frac{(n+k)!}{n!k!}\ln\left[\frac{(n+k)!}{n!k!}\right]
               \left(\frac{v}{c}\right)^{2k} .
\end{equation}

Let us go back to the wave function given in Eq.(\ref{cwf22a}).
As is illustrated in Figure~\ref{diracqm33}, its localization
property is dictated by the Gaussian factor which corresponds to
the ground-state wave function. For this reason, we expect that much
of the behavior of the density matrix or the entropy for the $n$-th
excited state will be the same as that for the ground state with
$n = 0.$  For this state, the density matrix and the entropy are
\begin{equation}\label{den33}
  \rho(z,z') = \left(\frac{1}{\pi \cosh(2\eta)}\right)^{1/2}
  \exp{\left\{-\frac{1}{4}\left[\frac{(z + z')^2}{\cosh(2\eta)}
                    + (z - z')^2\cosh(2\eta)\right]\right\}} ,
\end{equation}
and
\begin{equation}
  S = 2\left\{(\cosh\eta)^2 \ln(\cosh\eta) -
             (\sinh\eta)^2 \ln(\sinh\eta)\right\}  ,
\end{equation}
respectively.  The quark distribution $\rho(z,z)$ becomes
\begin{equation}
  \rho(z,z) = \left(\frac{1}{\pi \cosh(2\eta)}\right)^{1/2}
  \exp{\left(\frac{-z^2}{\cosh(2\eta)}\right) }.
\end{equation}

\end{widetext}

The width of the distribution becomes $\sqrt{\cosh\eta}$, and
becomes wide-spread as the hadronic speed increases.  Likewise,
the momentum distribution becomes wide-spread~\cite{knp86,hkn90pl}.
This simultaneous increase in the momentum and position
distribution widths is called the parton phenomenon in high-energy
physics~\cite{fey69a}.  The position-momentum uncertainty
becomes $\cosh\eta$.  This increase in uncertainty is due to our
ignorance about the physical but unmeasurable time-separation
variable.

Let us next examine how this ignorance will lead to the concept
of temperature.  For the Lorentz-boosted ground state with $n = 0$,
the density matrix of Eq.(\ref{den33}) becomes that of the harmonic
oscillator in a thermal equilibrium state if $(\tanh\eta)^2 $ is
identified as the Boltzmann factor~\cite{hkn90pl}.  For other
states, it is very difficult, if not impossible, to describe them
and thermal equilibrium states.  Unlike the case of temperature,
the entropy is clearly defined for all values of $n$.  Indeed,
the entropy in this case is derivable directly from the hadronic
speed.

The time-separation variable exists in the Lorentz-covariant world,
but we pretend not to know about it.  It thus is in Feynman's
rest of the universe.  If we do not measure this time-separation,
it becomes translated into the entropy.

\section*{CONCLUSIONS}

In this paper, we started with the Lorentz-invariant differential
equation of Feynman {\em et al.}~\cite{fkr71}.  This equation can
be separated into the Klein-Gordon equation for the free-flying
hadron and the harmonic-oscillator equation for the quarks inside
the hadron.  It was noted that there is a set of solutions
constituting a representation of the Poincar\'e group~\cite{knp86}.
While this set leads to many interesting consequences in
high-energy physics, it serves as the mathematical basis for
squeezed states in quantum optics.  This also serves as a
mathematical tool for illustrating Feynman's rest of the universe.

Starting from the physics of two-mode squeezed states, we were
able to give a physical interpretation to the time-separation
variable which is never mentioned in the present form of quantum
mechanics.

According to Feynman, the adventure of our science of physics is a
perpetual attempt to recognize that the different aspects of nature
are really different aspects of the same thing.   While this is his
interpretation of physics, the question is how to accomplish it.
One way is to prove that everything in physics comes from one
equation, as Newton did for classical mechanics.  Feynman's equation
of Eq.(\ref{diff11}) does not appear to generate all the physics, but
it could serve as the starting point in many branches of physics.

\end{document}